\documentclass[prx,twocolumn,superscriptaddress,citeautoscript,showpacs,amsart,longbibliography]{revtex4-2}

\usepackage{blindtext}
\usepackage{graphicx}
\usepackage{bm}
\usepackage{epsfig}
\usepackage{amssymb}
\usepackage{amsfonts}
\usepackage{braket}
\usepackage{color}
\usepackage{epstopdf}
\usepackage{hyperref}
\usepackage{float}
\usepackage{bibentry}
\usepackage{color}
\usepackage{multirow}
\usepackage[caption=false]{subfig}
\usepackage{amsfonts}
\usepackage{amsthm}
\usepackage{amsmath}
\usepackage{graphicx}
\usepackage{multirow}
\usepackage{color}
\usepackage{bbold}
\usepackage{bm}
\usepackage{times}
\usepackage{amsmath,bm,amsfonts}
\usepackage{dcolumn}
\usepackage{graphicx}
\usepackage{latexsym}
\usepackage[columnwise]{lineno}

\newcommand{\bpm}{\begin{pmatrix}}
\newcommand{\epm}{\end{pmatrix}}
\newcommand{\ba}{\begin{eqnarray}}
\newcommand{\ea}{\end{eqnarray}}
\newcommand{\bd}{\begin{displaymath}}

\graphicspath{{figures/}}

\begin{document}

\title{Superconductivity in Isolated Single Copper Oxygen Plane}

\author{Youngdo Kim}
\affiliation{Department of Physics and Astronomy, Seoul National University, Seoul 08826, Korea}

\author{Byeongjun Gil}
\affiliation{Department of Materials Science and Engineering and Research Instituteof Advanced Materials, Seoul National University, Seoul 08826, Korea}

\author{Sehoon Kim}
\affiliation{Department of Physics and Astronomy, Seoul National University, Seoul 08826, Korea}

\author{Yeonjae Lee}
\affiliation{Department of Physics and Astronomy, Seoul National University, Seoul 08826, Korea}

\author{Donghan Kim}
\affiliation{Department of Physics and Astronomy, Seoul National University, Seoul 08826, Korea}

\author{Jaeung Lee}
\affiliation{Department of Physics and Astronomy, Seoul National University, Seoul 08826, Korea}

\author{Jinyoung Kim}
\affiliation{Department of Physics and Astronomy, Seoul National University, Seoul 08826, Korea}

\author{Younsik Kim}
\affiliation{Department of Physics and Astronomy, Seoul National University, Seoul 08826, Korea}

\author{Miyoung Kim}
\affiliation{Department of Materials Science and Engineering and Research Instituteof Advanced Materials, Seoul National University, Seoul 08826, Korea}

\author{Changyoung Kim}
\email[Electronic address:$~$]{changyoung@snu.ac.kr}
\affiliation{Department of Physics and Astronomy, Seoul National University, Seoul 08826, Korea}

\date{\today}

\begin{abstract}
One of the central questions in cuprate superconductivity is if superconductivity can exist in an isolated single CuO$_2$ plane without any interlayer coupling. There have been numerous experimental efforts to answer this question, but it still has not been clearly resolved. Here we present a heterostructure system with an isolated half-unit-cell La$_{2-x}$Sr$_x$CuO$_4$ which has a single CuO$_2$ plane. Using \textit{in-situ} angle-resolved photoemission spectroscopy, we measured the electronic and gap structures of a single CuO$_2$ plane. We observed a \textit{d}-wave-like gap which closes somewhat above the bulk T$_c$. Moreover, almost identical gap properties are seen for both single CuO$_2$ plane and bulk. These observations lead us to the conclusion that the \textit{d}-wave superconductivity of cuprates also exists in a single CuO$_2$ plane. Our results demonstrate that cuprate superconductivity is essentially a two-dimensional phenomenon and provide a platform to study cuprate superconductivity in a purely two-dimensional system.

\end{abstract}
\maketitle

\section{Introduction}

High temperature superconductivity in cuprates occurs in CuO$_2$ planes and the T$_c$ generally increases with the number of CuO$_2$ planes. The latter fact led to the notion that the interlayer coupling may play a decisive role in the emergence of superconductivity in cuprates. Consequently, various models such as the interlayer tunneling model have been proposed to explain the role of interlayer coupling in cuprates~\cite{wheatley1988interlayer,anderson1995interlayer,leggett1996interlayer,moler1998images,anderson1998c}. However, none of these models has been proven successful. On the experimental side, it has not been clearly proven if the cuprate superconductivity can exist in a single CuO$_2$ plane. As a result, while most of the models assume the single CuO$_2$ plane superconductivity, it is still uncertain whether cuprate superconductivity is fundamentally a two-dimensional (2D) or three-dimensional phenomenon. 

The primary challenge in investigating the interlayer coupling effect lies in our inability to control it as an independent variable; the interlayer coupling is in general fixed (uncontrollable) for a given crystal structure. Moreover, the interlayer coupling, lattice structure, and superconducting properties vary in such a complex manner that it is difficult to extract the effect solely from the interlayer coupling. Therefore, the most appropriate way to isolate its effect is probably to eliminate it by making a system with a single CuO$_2$ plane. By comparing the properties of monolayer and bulk forms of the same material, one can identify the impact of interlayer coupling on the superconducting properties.

There have been attempts to realize superconductivity in ultrathin cuprate systems for decades~\cite{sato1997growth,rufenacht2003growth,bozovic1994superconducting,hetel2007quantum,zhong2016nodeless,ran2024ultrathin}, but none of them has reached the single CuO$_2$ plane limit. The core issue with the conventional approach is that the transport measurements which were used to detect the superconductivity may suffer from the connectivity issue for ultrathin films. As a result, transport measurements may incorrectly indicate insulating states for ultrathin films~\cite{sato1997growth,ran2024ultrathin,king2014atomic,kim2021capping}. The thinnest cuprate with superconductivity in transport measurements is an exfoliated Bi$_2$Sr$_2$CaCu$_2$O$_{8+\delta}$ monolayer containing two CuO$_2$ planes~\cite{jiang2014high,yu2019high}. For a single CuO$_2$ plane, not even a metallic transport result has been reported so far. 

On the other hand, angle resolved photoemission spectroscopy (ARPES) is immune from the connectivity issue as it does not require an order in the global scale. Especially, a measurement scheme to obtain the electronic structure of oxide monolayer systems was recently developed~\cite{sohn2021observation,kim2023heteroepitaxial,ko2023tuning}, which has been used to measure the electronic structure of cuprate monolayers~\cite{kim2023growth}. In this study, we utilized the newly developed \textit{in-situ} ARPES method to obtain the electronic structures of a high quality isolated single CuO$_2$ plane system without any neighboring CuO$_2$ planes. The high resolution electronic structure data allowed us to observed a $d$-wave superconducting gap, demonstrating for the first time that superconductivity exist in a single CuO$_2$ plane. Our results show that superconductivity seen in bulk cuprates is fundamentally a 2D phenomenon.

\begin{figure*}[]
	\textsc{}	\includegraphics[width=\linewidth]{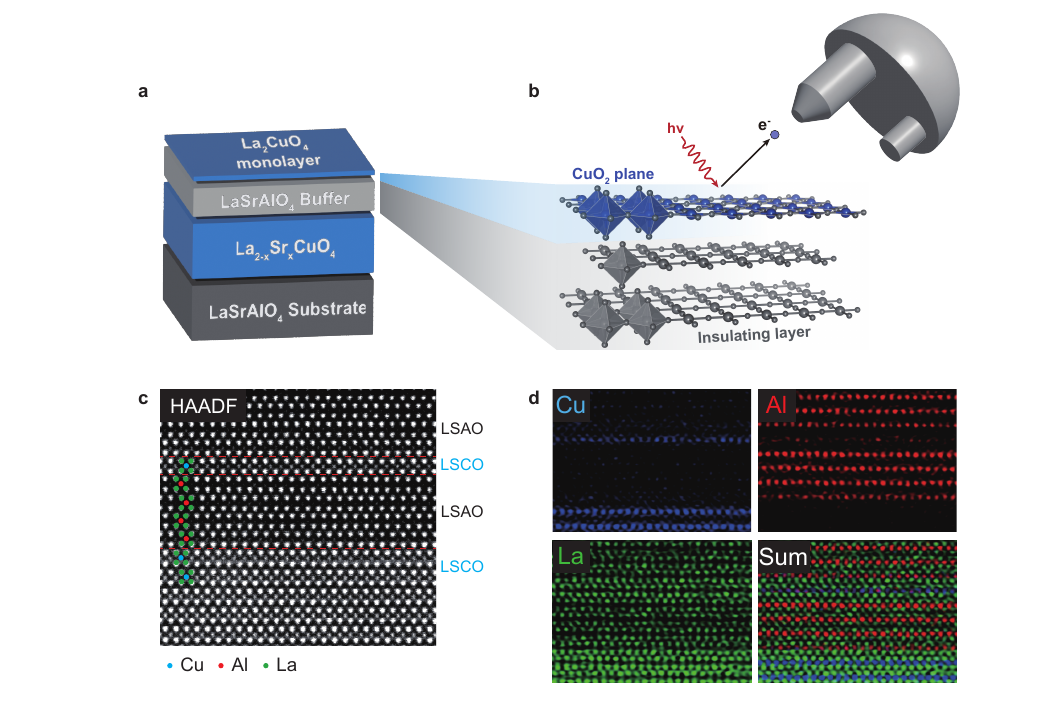}
	\caption{\textbf{Construction of single CuO$_2$ plane system.} A schematic illustration of (a) the heterostructure with single CuO$_2$ plane used in the study and (b) ARPES measurements on isolated single CuO$_2$ plane. (c) High-angle annular dark field scanning transmission electron microscopy (HAADF-STEM) and (d) energy-dispersive X-ray spectroscopy (EDX) images of the heterostructure.}
	\label{fig:1}
\end{figure*}

\section{Growth of LSCO monolayers}

ARPES measurements of a monolayer system grown on an insulating substrate can suffer from the charging effect and thus require a heterostructure to compensate for the escaping photoelectrons~\cite{sohn2021observation}. The heterostructure consists of a monolayer, a conducting layer, an insulating layer and a substrate. The conducting layer provides a macroscopic conducting path to the monolayer system while the insulating layer decouples the electronic structures of the monolayer and the conducting layer.

We chose La$_{2-x}$Sr$_x$CuO$_4$ (LSCO) as the target material. LSCO has a single CuO$_2$ plane in each layer and has a Ruddlesden-Popper structure with $n=1$ which is suitable for heterostructure engineering. LSCO was also chosen for the conducting layer for structural stability. LaSrAlO$_4$ (LSAO) was used for the insulating layer and the substrate which is the most used substrate for LSCO film growth. This also ensures that the heterostructure is all in the same Ruddlesden-Popper phase with $n=1$, resulting in a high-quality monolayer system.

When building a monolayer system with doped materials, the effects of cation intermixing should be considered~\cite{logvenov2009high,smadici2009superconducting,gozar2008high,yacoby2013atomic,kim2023growth}. In our monolayer system, cation intermixing occurs between the topmost LSCO monolayer and the LSAO buffer layer. Since the La to Sr ratio in the LSAO buffer layer is 1:1, the LSCO monolayer can be too heavily doped with diffused Sr. Therefore, La$_2$CuO$_4$ (LCO) with no Sr doping was used as the topmost monolayer to make the hole doping of the single CuO$_2$ plane lie in the superconducting dome. A schematic for the resulting heterostructure is shown in Fig. 1(a). Figure 1(b) shows a schematic of ARPES measurement in our LCO monolayer system. The ARPES measurement of this system measures the electronic structure of the topmost LCO monolayer and the LSAO buffer layer directly below it. Since insulating LSAO has no spectral weight near the Fermi level, only the electronic structure of the LCO monolayer is visible near the Fermi level.

In order to confirm the quality of our heterostructure, atomic visualization of the LSCO monolayer system was obtained by using scanning transmission electron microscopy (STEM). An LSAO capping layer was deposited on top of the monolayer for STEM measurement. Figure 1(c) shows a high-angle annular dark field STEM (HAADF-STEM) image of the LCO monolayer heterostructure. From the top, the LSAO capping layer, LCO monolayer, LSAO buffer layer, and LSCO conducting layer are well resolved. It is also visible that all layers are well-aligned along the layered perovskite structure. Low magnification HAADF-STEM image (Supplementary Fig. 1) shows a high-quality heterostructure without disorder over a large area. 

The single CuO$_2$ plane in the LCO monolayer system is more clearly observed in an energy dispersive X-ray spectroscopy (EDX) image. Figure 1(d) shows the distribution of each element in the heterostructure. Note that elements La, Cu, and Al are all in the proper positions in the heterostructure. Particularly, the single layer of Cu atom is evident between the neighboring Al layers, indicating that an epitaxial and high quality single CuO$_2$ plane was grown.

\begin{figure*}[]
	\textsc{}	\includegraphics[width=\linewidth]{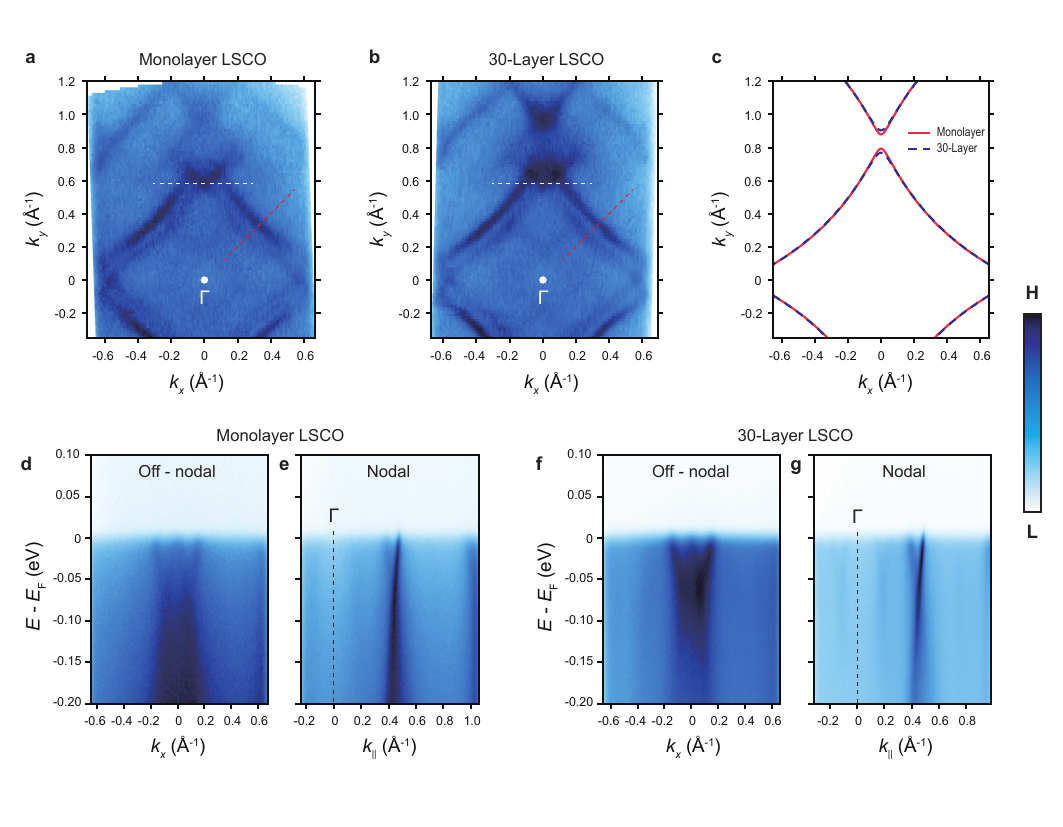}
	\caption{\textbf{Electronic structure of monolayer and 30-layer La$_{2-x}$Sr$_x$CuO$_4$.} Fermi surface maps of (a) monolayer and (b) 30 layer (15 UC) of LSCO with LSAO buffer layer. (c) Fermi surface of a monolayer and 30-layer LSCO fitted with tight-binding model. (d-e) Off-nodal and nodal cuts of a monolayer along the white and red dashed lines in (a), respectively. (f-g) Off-nodal and nodal cuts of 30-layer LSCO along the white and red dashed lines in (b), respectively.}
	\label{fig:2}
\end{figure*}

\section{Electronic structure measurement}

Since the monolayer of (nominal) LCO was hole doped due to the Sr intermixing with the LSAO buffer layer, we will refer it as the monolayer LSCO from now on to avoid potential confusion. The electronic structure of the monolayer LSCO measured by ARPES is summarized in Fig. 2. Figure 2(a) shows a Fermi surface map of the monolayer LSCO. At first glance, the Fermi surface looks quite different from the known Fermi surface, with multiple folding bands. However, these folding bands also appear in the electronic structure of thick films, shown in Fig. 2(b). Moreover, the low energy electron diffraction image of the thick film (Supplementary Fig. 2(a)) indicates a $4\times4$ structure which is consistent with the folding bands (Supplementary Fig. 2(b)). Then, based on the $4\times4$ reconstruction picture, we can identify the original band which is found to be quite similar to that of the bulk LSCO~\cite{ino2002doping,yoshida2006systematic,razzoli2010fermi,zhong2022differentiated}. 

Once the original band is identified, we can determine the change in hole doping due to Sr intermixing by fitting the Fermi surface with a single band tight-binding model~\cite{yoshida2006systematic,pavarini2001band,yoshida2007low,kim2023growth}.
\begin{equation*}
	\begin{aligned}
		\epsilon_{k} = & \epsilon_{0} - 2t[\cos (k_{x}a) + \cos (k_{y}a)]- 4t^{\prime}\cos (k_{x}a)\cos (k_{y}a)\\
		& - 2t^{\prime\prime}[\cos (2k_{x}a) + \cos (2k_{y}a)]
	\end{aligned}
\end{equation*}
Here, $t$, $t^\prime$, and $t^{\prime\prime}$ represent the first, second, and third nearest-neighbor hoppings, respectively. We calculated the Fermi surface area from the fitted band, from which the hole doping can be estimated based on the Luttinger theorem\cite{yoshida2006systematic}. So, obtained hole doping level is 0.26, showing that the change in hole doping due to intermixing is quite dramatic.

\begin{figure*}[]
	\textsc{}	\includegraphics[width=\linewidth]{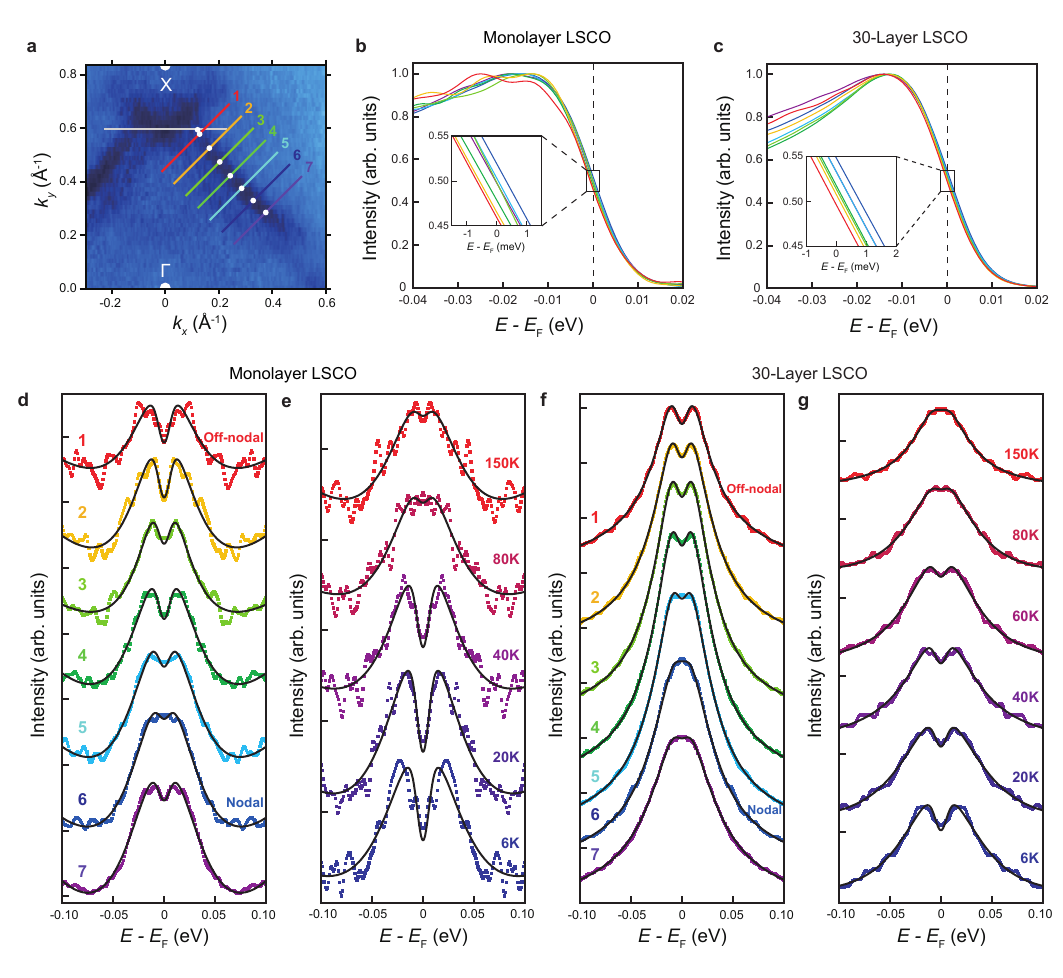}
	\caption{\textbf{Superconducting gap analysis.} (a) Fermi surface of LSCO monolayer. White dots indicate the Fermi momenta \textit{k$_F$} where \textit{k$_F$} energy distribution curves (EDCs) are taken. (b) EDCs at \textit{k$_F$}'s in (a), with each color corresponding to the line color in (a). The inset zooms in near the Fermi level to show the leading-edge shift. (c) EDCs from a 30-layer LSCO obtained with the same method as in the monolayer LSCO case. (d-g) Symmetrized EDCs of monolayer and 30-layer LSCO. Black lines are fitting results. (d) Angle-dependent symmetrized EDCs of the monolayer LSCO. (e) Temperature-dependent symmetrized EDCs of monolayer LSCO. EDCs are extracted at \textit{k$_F$} of off-nodal cut (gray line in (a)). (f-g) Angle- and temperature-dependent symmetrized EDCs of 30-layer LSCO, respectively. All EDCs were processed in the same way as in case of the monolayer LSCO. }
	\label{fig:3}
\end{figure*}

For the comparison between monolayer and bulk LSCO, a 30-layer LSCO with a Sr doping of $x\approx 0.25$ was grown on an LSAO substrate. The Fermi surface of 30-layer LSCO shown in Fig. 2(b) is almost indistinguishable from that of the monolayer; the two are nearly identical, including the folded bands. The Fermi surface of the bulk LSCO is also fitted with a tight binding model. The fitted Fermi surfaces of the monolayer and bulk LSCO are plotted in Fig. 2(c). The hole doping of 30-layer LSCO was determined to be 0.27. The off-nodal and nodal cuts for both the monolayer and the 30-layer LSCO are plotted in the figure 2 (d-g). The features of the folded bands are well resolved in both off-nodal and nodal cuts. The only difference between the two is in the spectral intensity; the spectra from the bulk are understandably more intense. Otherwise, they have almost identical characteristics.

\section{Superconducting gap structure}

\begin{figure*}[t!]
	\textsc{}	\includegraphics[width=\linewidth]{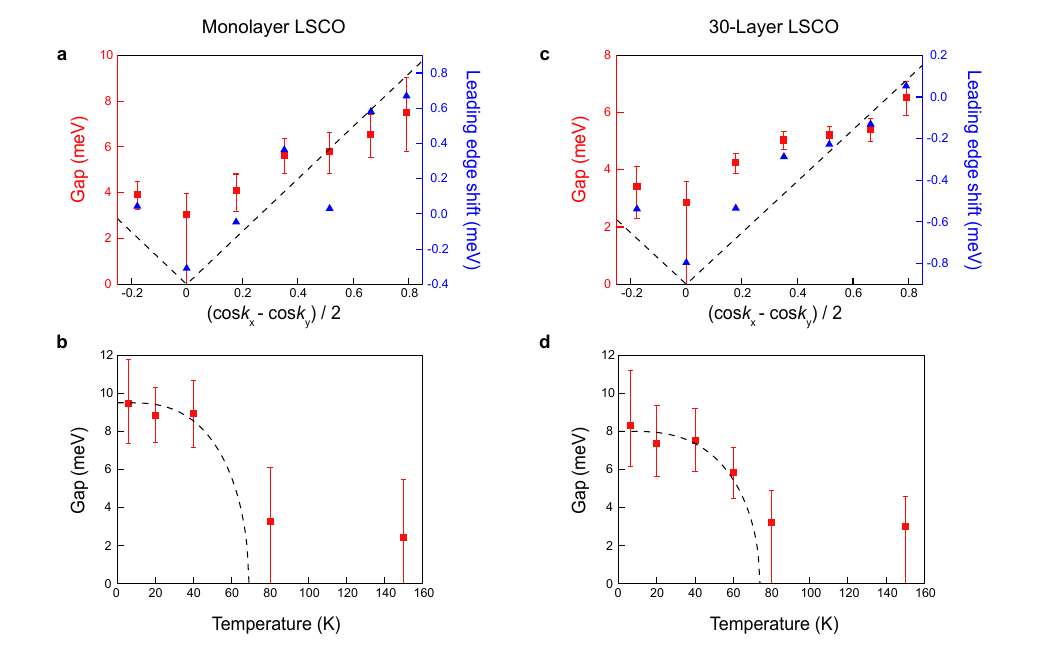}
	\caption{\textbf{Momentum- and temperature-dependent gap behavior.} Red squares and blue triangles represent gap behavior determined by fitting and leading-edge shifts, respectively. Black dashed lines are guides for the d-wave gap (momentum-dependence) and BCS gap function (temperature-dependence). (a) Momentum- and (b) temperature-dependent gaps for the monolayer. (c) Momentum- and (d) temperature-dependent gaps for the 30-layer LSCO. }
	\label{fig:4}
\end{figure*}

Resistivity measurements show that our 30-layer LSCO is a superconductor with a T$_c$ of 35 K (Supplementary Fig. 3). Therefore, we can expect that the monolayer system with a similar doping may possess superconductivity. In order to investigate the possible superconductivity in the single CuO$_2$ plane sample and its nature, we analyzed the momentum- and temperature-dependent gap behavior of the monolayer and compared it with that of the 30-layer sample. As depicted in Fig. 3(a), ARPES spectra were taken along the indicated cuts from the node to antinode. Figure 3(b) shows energy distribution curves (EDCs) at Fermi momenta of cuts in Fig. 3(a). As the momentum changes from the node to antinode, the leading edge of spectra moves towards the higher binding energy side below the Fermi level, showing the typical behavior of a $d$-wave superconducting gap. The EDCs of the 30-layer LSCO were also measured in the same way and plotted in Figure 3(c), exhibiting the same characteristics as the monolayer LSCO.

We also symmetrized the EDCs of the monolayer about the Fermi level and plot them in Fig. 3(d). The symmetrized EDCs show the absence of a gap at the nodal point and a gap opening towards the anti-nodal point. Temperature-dependent gap measurements were also performed at the anti-nodal cut. The symmetrized temperature dependent EDCs are plotted in Fig. 3(e). The anti-nodal gap closes at a temperature between 40-80 K as the temperature increases, which is significantly higher than the T$_c$ of LSCO with a hole doping $x\approx 0.26$. The EDCs of the 30-layer LSCO symmetrized in the same way are shown in Figs. 3(f) and 3(g). The 30-layer LSCO also exhibits a $d$-wave gap characteristics and gap closing between 60-80 K. The similar gap behaviors between the monolayer and 30-layer systems are strongly suggestive of the superconducting nature of the gap observed in the monolayer system.

For a quantitative analysis of the superconducting gap, the symmetrized EDCs were fitted with a Norman function~\cite{norman1998phenomenology,wu2024nodal}
\begin{equation*}
	\begin{aligned}
		A(\omega) = A_{0} \operatorname{Im}(1/(\omega - \Sigma(\omega)))
	\end{aligned}
\end{equation*}
with the self-energy term expressed as
\begin{equation*}
	\begin{aligned}
		\Sigma(\omega) = - i \Gamma +\Delta^{2}/\omega
	\end{aligned}
\end{equation*}
Here, $\Gamma$ is the lifetime of quasiparticles and $\Delta$ is the gap. The fitting results are plotted as black curves in Figs. 3(d-g), demonstrating a good fit to the symmetrized EDC.

The gap sizes obtained from the fitting are shown in Fig. 4 along with the leading-edge shift values extracted from Figs. 3(b) and 3(c). Figure 4(a) plots the momentum-dependent gap size and leading-edge shift against the $d$-wave parameter $(\cos (k_{x}) - \cos (k_{y}))/2$. Both gap size and leading-edge shift have a minimum value at the nodal point where the $d$-wave parameter goes to zero, and increase with the $d$-wave parameter toward the anti-nodal point.

The temperature dependent gap size plotted in Fig. 4(b) exhibits gap closing at a temperature between 40-80 K. Note that the fitted gap size at 6K is slightly larger than the maximum value for the angle-dependent gap data. Considering the fact that the temperature-dependent data was taken at a momentum point to the anti-nodal point than the measurement range of the momentum-dependent data, this also supports the $d$-wave gap symmetry of the monolayer LSCO. The momentum- and temperature-dependent gap size of the 30-layer LSCO is also shown in Figs. 4(c) and 4(d). The 30-layer LSCO also shows a very similar $d$-wave gap symmetry behavior as that of the monolayer. In addition, the overall gap size and gap closing temperature are also similar.

A point to note is that, although the symmetrized EDCs of the nodal point and high-temperature data shows no signs of the gap, their fitting results indicate a non-vanishing value. This is from an artifact of the fitting. Due to the characteristics of the fitting function and instrumental resolution, the symmetrized EDCs exhibit nearly identical shapes in the region where the gap is sufficiently small (0-3 meV). In this region, the fitted gap size shows a finite value with the error bar extending to the zero value (Supplementary Fig. 4 and Fig. 5).

\section{Discussion}

As we have closely investigated the gap nature of the monolayer LSCO, it is necessary to address if the gaps in both systems have a superconducting origin. We argue that the similarity in electronic structures and gap characteristics between the monolayer and 30-layer LSCO are indications that their physical properties, particularly the superconducting ones, are likely to be the same. Both exhibit typical d-wave symmetry of cuprate superconductors and have a maximum value of around 10 meV, similar to the typical superconducting gap value for bulk LSCO~\cite{shi2008coherent,terashima2007anomalous,zhong2022differentiated}.

One peculiar aspect of the gap is that both gaps appear to close around 60-80 K in the temperature dependent data, higher than the T$_c$ of the 30-layer sample determined from the resistivity. Given that the samples are in the over-doped region, the gaps are unlikely to be pseudogaps. On the other hand, the presence of a superconducting gap at temperatures above T$_c$ has been widely reported not only in LSCO~\cite{yoshida2009universal} but also in other cuprates~\cite{kondo2011disentangling,kondo2015point,chen2019incoherent,he2021superconducting}. In most cases, it is attributed to preformed pairs~\cite{kondo2011disentangling,kondo2015point} or superconducting fluctuations~\cite{chen2019incoherent,he2021superconducting} that persist above T$_c$. 

Considering all aspects discussed above, the gap shared by the monolayer and 30-layer LSCO is very likely a superconducting gap. We therefore conclude that the monolayer and 30-layer LSCO exhibit the same superconducting properties. Our results demonstrate that the superconductivity of cuprates, as we know it, exists in an LSCO monolayer or a single CuO$_2$ plane system. 

This is the first observation of superconductivity in a single CuO$_2$ plane without neighboring CuO$_2$ planes. Our results indicate that cuprate superconductivity is a 2D phenomenon, and interlayer coupling is not essential for the existence of the superconductivity. On the other hand, while no significant difference is observed between the monolayer and bulk cuprate system in this case, we may not exclude the possibility that the interlayer coupling could still play a role in the dimensionality effect in cuprates.

For broader studies on the dimensionality effect in cuprate systems, it is essential to control the carrier doping of the single CuO$_2$ plane. In current monolayer system using LCO and LSAO buffer layer, we were unable to obtain a lower doping than the current doping ($x\approx 0.26$) due to the Sr intermixing. Controlling the doping of a monolayer can be achieved by replacing the buffer layer, chemical doping of the monolayer, or alkali metal dosing. With further doping control, we can expect that various phenomena appear in the phase diagram such as pseudogaps or charge ordering and that they can be studied in the monolayer system, broadening our understanding of cuprate superconductivity.

\section*{Acknowledgements}
This work is supported by the National Research Foundation of Korea(NRF) grant funded by the Korea government(MSIT)(No. 2022R1A3B1077234) and Global Research Development Center (GRDC) Cooperative Hub Program through the National Research Foundation of Korea (NRF) funded by the Ministry of Science and ICT (MSIT)(Grant No. RS-2023-00258359). STEM measurement was supported by the National Research Foundation of Korea (NRF) grant funded by the Korean government (MSIT) (NRF-2022R1A2C3007807).

\bibliographystyle{apsrev}
\bibliography{bib_SinglelayerSC.bib}

\end{document}